\documentstyle[aps,prl,epsf,preprint]{revtex}

\begin{document}
\draft
\title{Specific Heat Study of the Field-Induced Magnetic Ordering in the Spin Gap System TlCuCl$_3$}

\author{A. Oosawa, H. Aruga Katori$^1$ and H. Tanaka}

\address{Department of Physics, Tokyo Institute of Technology, Oh-okayama, Meguro-ku, Tokyo 152-8551\\
$^1$RIKEN (The Institute of Physical and Chemical Research), Wako, Saitama 351-0198}

\date{\today}

\maketitle

\begin{abstract}
Specific heat measurements have been performed in the coupled dimer system TlCuCl$_3$, which has a singlet ground state with the excitation gap $\Delta \simeq 7.5$K. The cusplike anomaly indicative of the 3D magnetic ordering was clearly observed in magnetic fields higher than the critical field $H_{\rm g}$ corresponding to the gap $\Delta$. The phase boundary determined by the present specific heat measurements coincides with that determined by previous magnetization measurements. The phase boundary can be described by the power law $\left[ H_{\rm c}(T)-H_{\rm g} \right] \propto T^{\phi}$ with $\phi=2.1(1)$. This result supports the magnon Bose condensation picture for the field-induced magnetic ordering in TlCuCl$_3$ [Nikuni {\it et al}., Phys. Rev. Lett. {\bf 84} (2000) 5868].
\end{abstract}


Recently, quantum spin systems such as $S=1/2$ two-leg spin ladders, $S=1/2$ ferromagnetic-antiferromagnetic alternating chains and $S=1$ antiferromagnetic chains have been the focus of much interest for both theorists and experimenters, because these systems have the singlet ground state with the excitation gap $\Delta$. \par
One-dimensional gapless spin systems do not undergo three-dimensional (3D) magnetic ordering. However, real quasi-1D gapless spin systems can undergo 3D ordering at low temperatures due to finite 3D exchange interactions. For example, the $S=5/2$ quasi-1D spin system (CH$_3$)$_4$NMnCl$_3$ (abbreviated TMMC) undergoes 3D magnetic ordering at $T_{\rm N}=0.835$K at zero field \cite{Takeda}. \par
On the contrary, the above mentioned spin gap systems remain paramagnetic down to $T=0$ K, even if there are weak 3D exchange interactions. However, when a magnetic field higher than the critical field $H_{\rm g}=\Delta/g\mu_{\rm B}$ is applied, the ground state becomes gapless, so the systems can undergo 3D magnetic ordering with the help of 3D exchange interactions. Such field-induced 3D ordering was observed in some quasi-1D spin gap systems, {\it e.g.}, the $S=1/2$ Heisenberg ladder system Cu$_2$(C$_5$H$_{12}$N$_2$)$_2$Cl$_4$ \cite{Hammer1,Hammer2,Chaboussant,Hagiwara}, the $S=1$ antiferromagnetic chain systems Ni(C$_5$H$_{14}$N$_2$)$_2$N$_3$(ClO$_4$) \cite{Honda1} and Ni(C$_5$H$_{14}$N$_2$)$_2$N$_3$(PF$_6$) \cite{Honda2}, and the $S=1/2$ exchange alternating system (CH$_3$)$_2$CHNH$_3$CuCl$_3$ \cite{Manaka}. \par
Recently, we observed the field-induced 3D magnetic ordering in the spin gap system TlCuCl$_3$ by means of magnetization measurements \cite{Oosawa}. This compound is isomorphous with KCuCl$_3$ which belongs to the monoclinic space group $P2_1/c$ \cite{Willett,Oosawa2}. TlCuCl$_3$ contains planar dimers of Cu$_2$Cl$_6$, in which Cu$^{2+}$ ions have spin-$1/2$. These dimers are stacked on top of one another to form infinite chains parallel to the crystallographic $a$-axis. These chains are located at the corners and centre of the unit cell in the $bc$-plane, and are separated by Tl$^+$ ions. \par
From previous magnetic measurements, the ground state of TlCuCl$_3$ was found to be a spin singlet with the excitation gap $\Delta \simeq 7.5$K \cite{Oosawa,Takatsu,Shiramura,Tanaka}. The origin of the gap for TlCuCl$_3$ is the strong antiferromagnetic exchange interaction in the chemical dimers of Cu$_2$Cl$_6$, similar to KCuCl$_3$ \cite{Oosawa2,Kato1,Kato2,Cavadini,Kato3}. The magnetic dimers are coupled three-dimensionally through interdimer interactions. \par
TlCuCl$_3$ undergoes 3D magnetic ordering in magnetic fields higher than $H_{\rm g}\sim 6$ T. The feature of the magnetization at low temperatures, including the phase transition temperature $T_{\rm N}$, and the field dependence of $T_{\rm N}$ are summarised as follows. The magnetization exhibits the cusplike minimum at $T_{\rm N}$. The phase boundary on the temperature vs field diagram is independent of the field direction, when normalized by the $g$-factor, and can be represented by the power law
$$\left( g/2 \right) \left[H_{\rm N}(T)-H_{\rm g}\right] \propto T^{\phi} , \eqno(1)$$
with $\phi=2.2$, where $H_{\rm N}(T)$ is the transition field at temperature $T$. \par
These features for the magnetization and the phase boundary have been described in terms of the Bose-Einstein condensation (BEC) of excited triplets (magnons) \cite{Nikuni}. However, there is some ambiguity for determining $T_{\rm N}$ from the magnetization data, because the cusplike minimum of the magnetization near $T_{\rm N}$ is not sharp, but rounded \cite{minimum}. In order to obtain conclusive evidence of the field-induced magnetic ordering, we performed specific heat measurements, and report the results in this paper. \par


Single crystals of TlCuCl$_3$ were grown from a melt by the Bridgman method. The details of preparation have been reported in reference 9. The crystal is cleaved along the (0,1,0) and (1,0,${\bar 2}$) planes, which are perpendicular to each other. \par 
The specific heat measurements of single crystals of TlCuCl$_3$ were carried out at temperatures down to 0.6K in magnetic fields up to 12T, using a Mag Lab$^{\rm HC}$ microcalorimeter (Oxford Instruments) in which the relaxation method was employed. In this paper, the magnetic field was applied perpendicular to the (0,1,0) and (1,0,${\bar 2}$) planes. 

\begin{figure}[ht]
\vspace*{5cm}
\begin{minipage}{7.5cm}
 \epsfxsize=75mm
  \centerline{\epsfbox{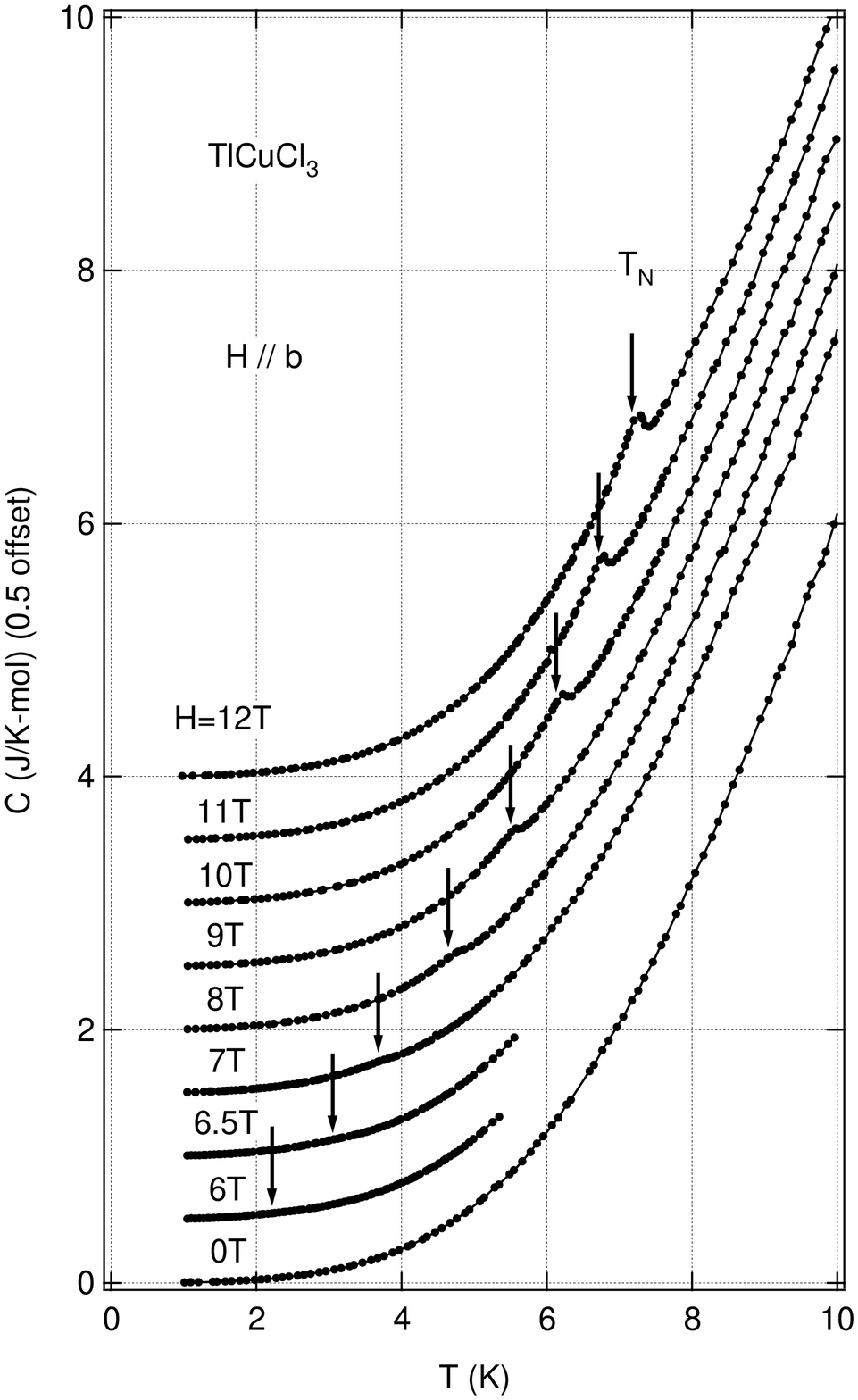}}
\begin{center}
(a)
\end{center}
\end{minipage}
\begin{minipage}{7.5cm}
 \epsfxsize=75mm
  \centerline{\epsfbox{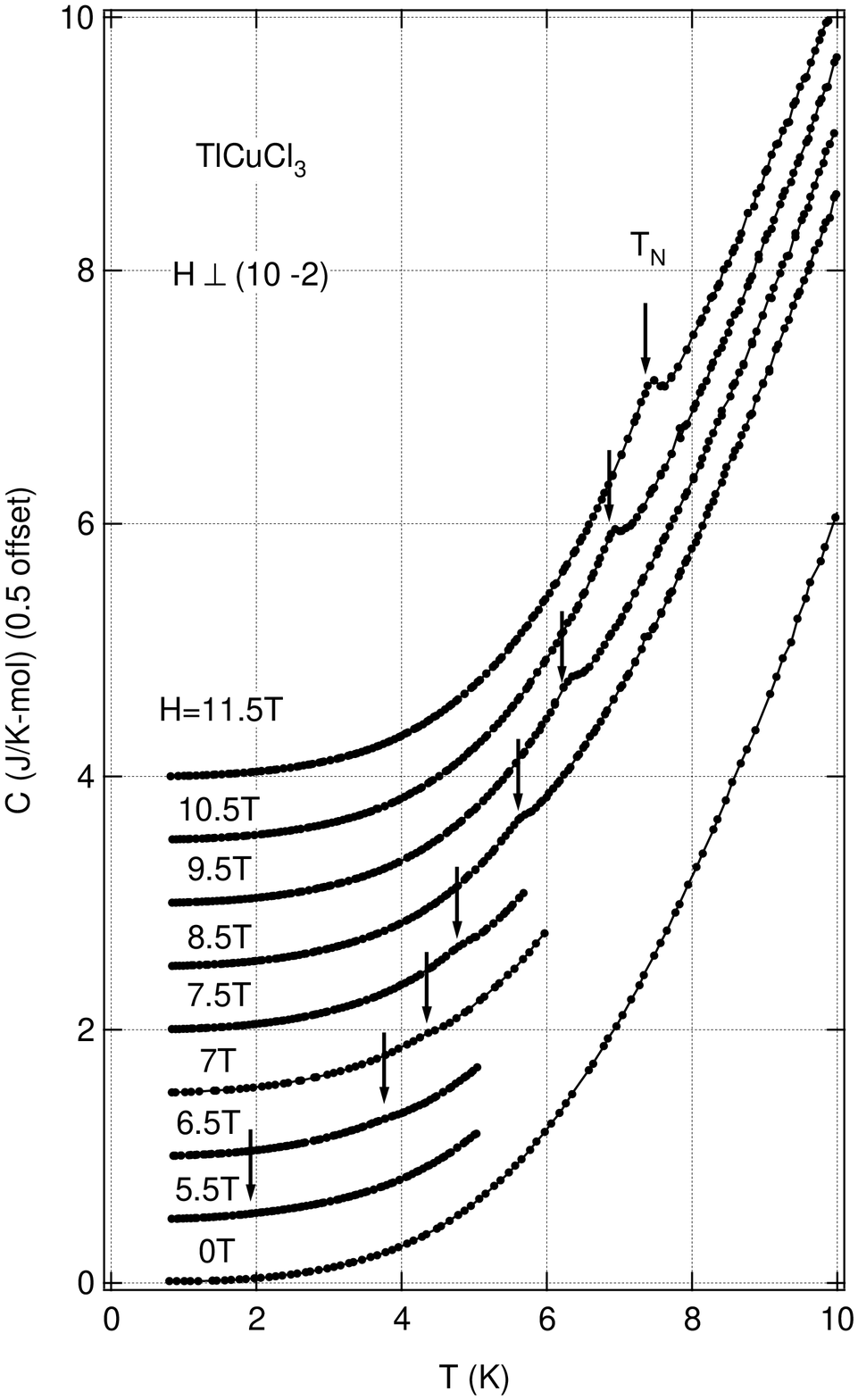}}
\begin{center}
(b)
\end{center}
\end{minipage}
\vspace*{1cm}
	\caption{The temperature dependence of the specific heat in TlCuCl$_3$ at various magnetic fields for (a) $H\parallel b$ and (b) $H\perp (1,0,{\bar 2})$.}
	\label{Tscan}
\end{figure}

Figure \ref{Tscan} shows the temperature dependence of the total specific heat in TlCuCl$_3$ measured at various magnetic fields for $H \parallel b$ and $H \perp (1,0,{\bar 2})$. For $H=0$ T, no
anomaly is seen down to 0.8 K, as expected from the gapped ground state. \par
When the applied field is higher than the critical field $H_{\rm g}\approx$ 5.4T for $H\parallel b$ and $\sim5.0$T for $H\perp (1,0,{\bar 2})$) corresponding to the energy gap $\Delta$ $(\sim 7.5$K), a small cusplike anomaly indicative of the magnetic ordering is observed.  The difference in the critical fields in the two field directions is due to the anisotropy of the $g$-factor \cite{Oosawa}. \par
The difference between the specific heats for $H=0$ T and for $H>H_{\rm g}$ is slight except around the N\'{e}el temperature $T_{\rm N}$, although large increases of the magnetic specific heat may be expected for $H>H_{\rm g}$. We infer that this is because the induced magnetic moment is small for $H\sim H_{\rm g}$, since the saturation field $H_{\rm s}$ is $\sim 100$ T, which is more than ten times as large as $H_{\rm g}$ \cite{Shiramura,Tatani}. \par
\begin{figure}[ht]
\vspace*{5cm}
\begin{minipage}{7.5cm}
 \epsfxsize=75mm
  \centerline{\epsfbox{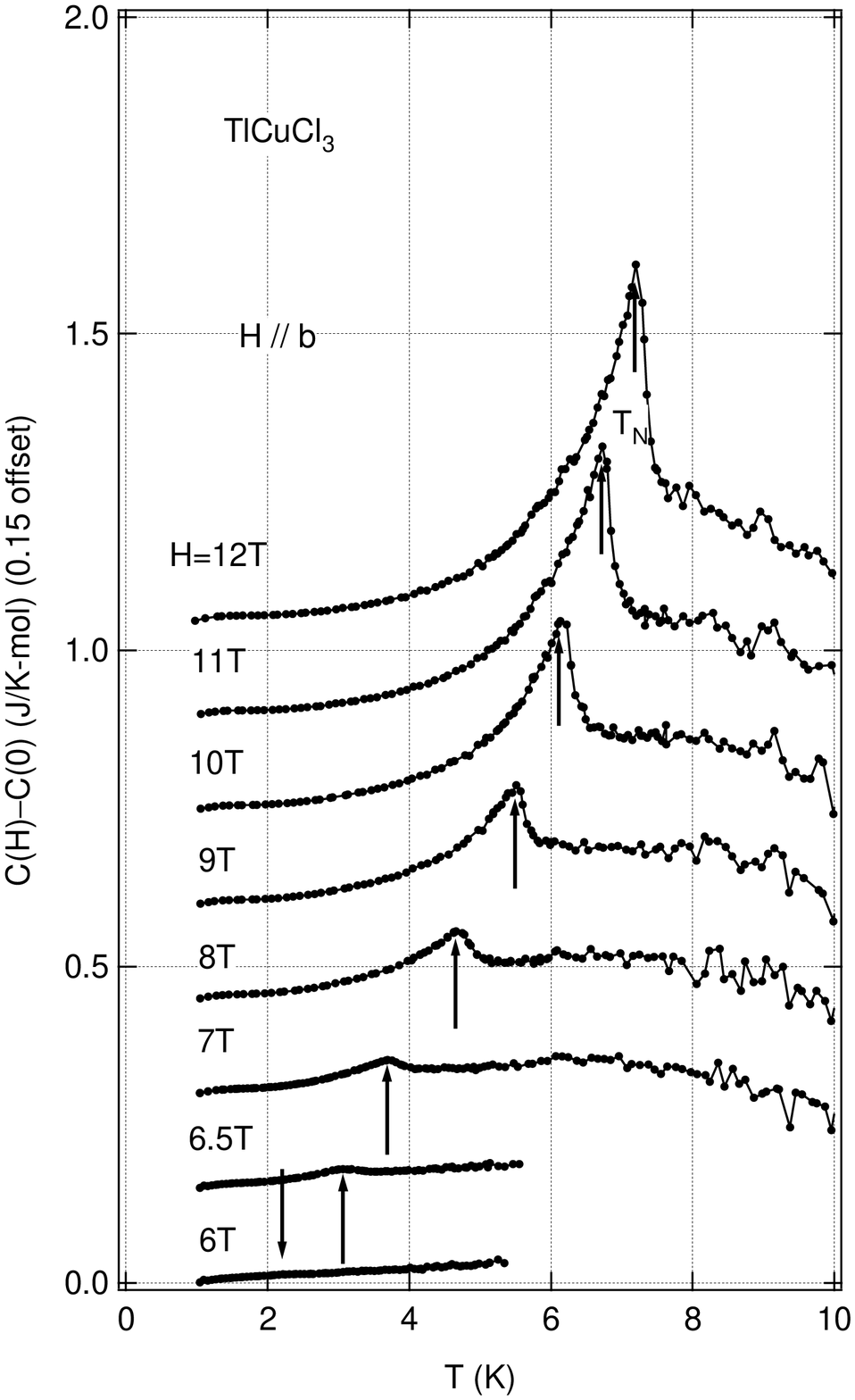}}
\begin{center}
(a)
\end{center}
\end{minipage}
\begin{minipage}{7.5cm}
 \epsfxsize=75mm
  \centerline{\epsfbox{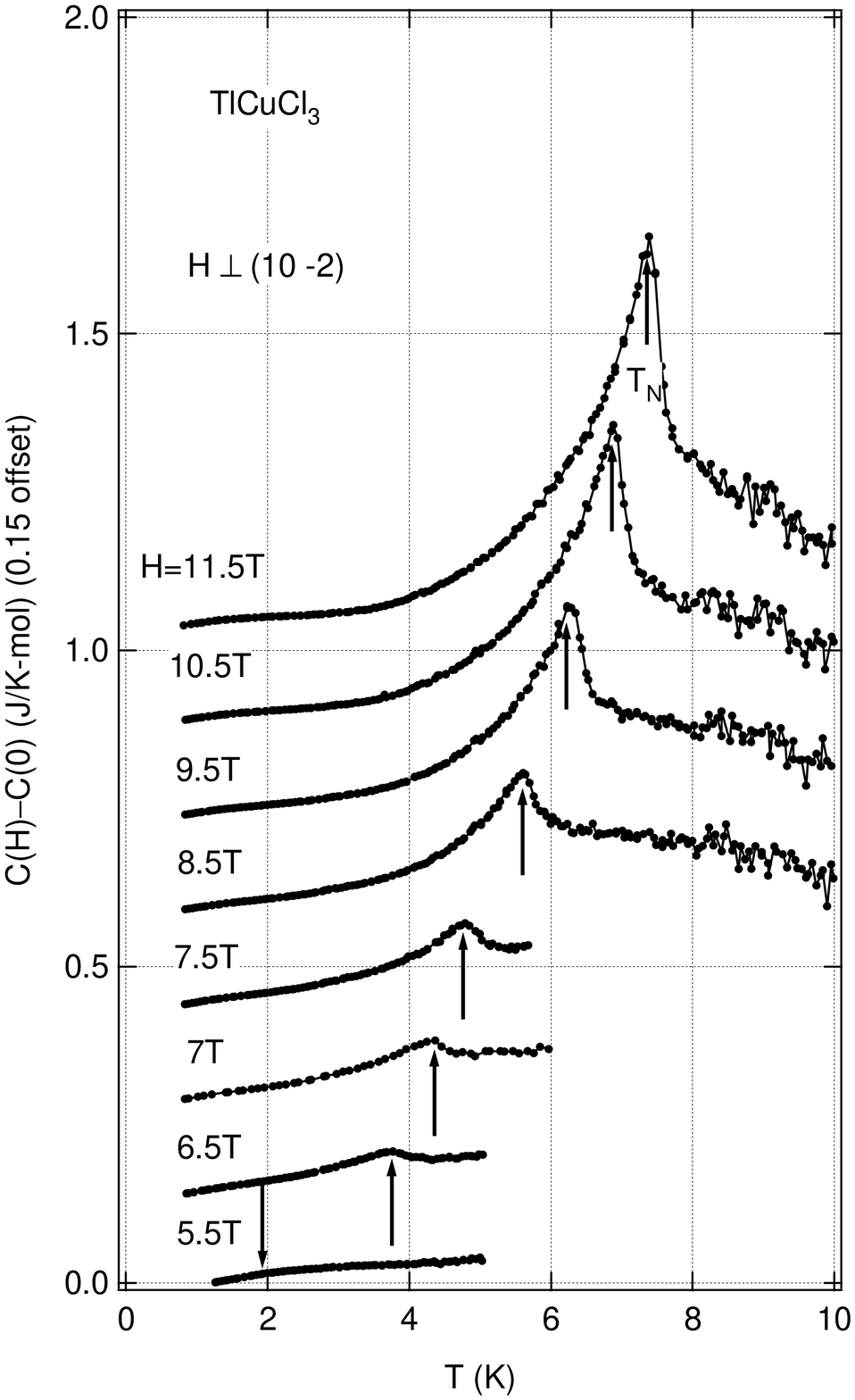}}
\begin{center}
(b)
\end{center}
\end{minipage}
\vspace*{1cm}
	\caption{The temperature dependence of the difference between the specific heat $C(H)$ at magnetic field $H$ and $C(0)$ at zero field.}
	\label{Tscan2}
\end{figure}

In Fig. \ref{Tscan2}, we plot the difference between the specific heat $C(H)$ for a magnetic field $H$ and $C(0)$ for $H=0$ as a function of temperature. $C(H)-C(0)$ can be regarded as the specific heat relevant to the field-induced magnetic ordering. However, it is noted that $C(H)-C(0)$ does not express the magnetic specific heat exactly, since $C(0)$ includes both the lattice part ($\propto T^3$) and the spin-gap part ( $\propto \exp(-\Delta/T)$ ). The $\lambda$-like anomaly due to the phase transition is clearly observed in the plot of $C(H)-C(0)$ vs $T$. With decreasing magnetic field, the transition temperature decreases, and the $\lambda$-like anomaly becomes smaller. This means that the spin moment relevant to the ordering is reduced when the magnetic field reaches $H_{\rm g}$. For $H<6.5$ T, it is difficult to determine the transition temperature $T_{\rm N}$. \par
\begin{figure}[ht]
\vspace*{5cm}
\begin{minipage}{7.5cm}
 \epsfxsize=75mm
  \centerline{\epsfbox{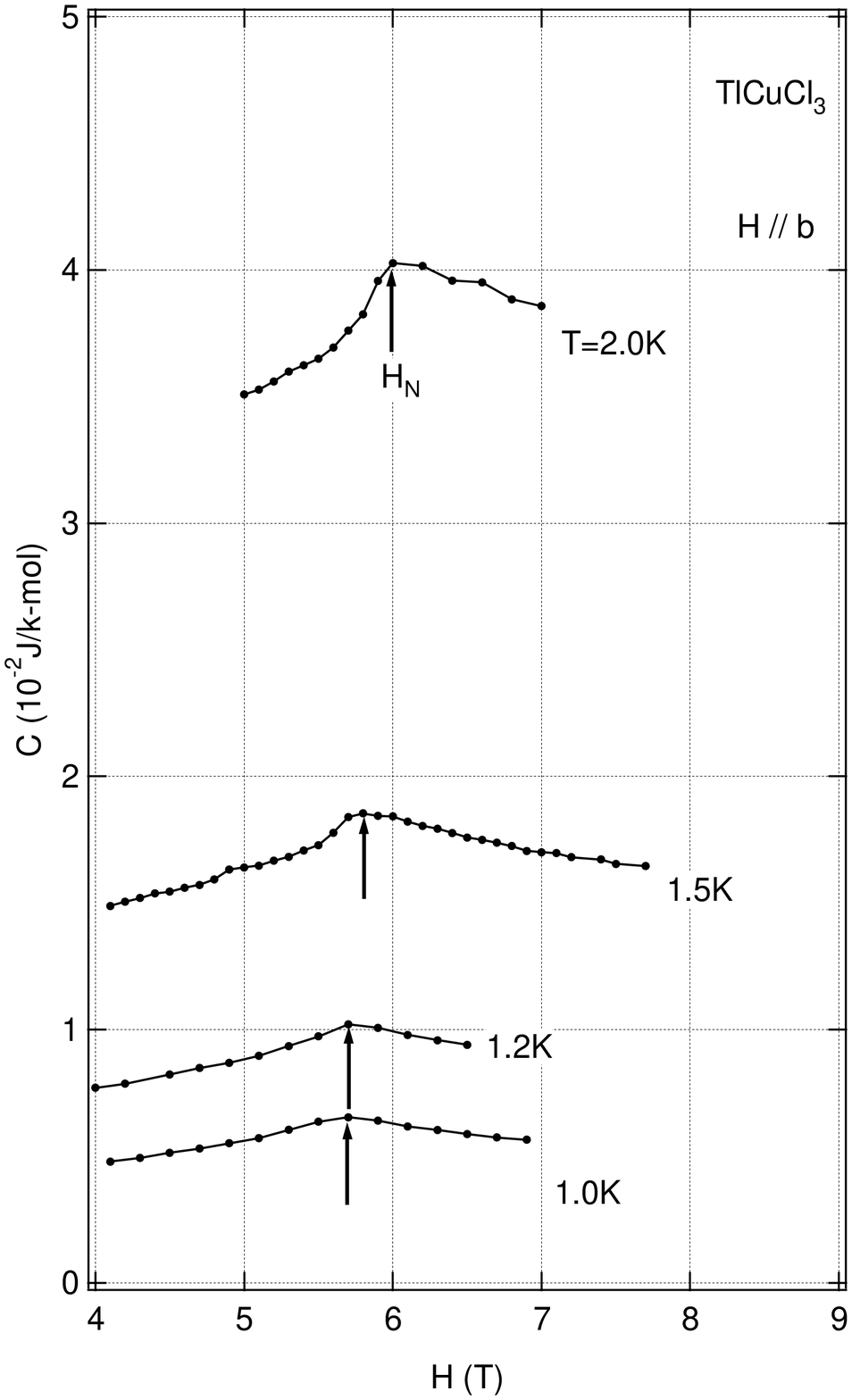}}
\begin{center}
(a)
\end{center}
\end{minipage}
\begin{minipage}{7.5cm}
 \epsfxsize=75mm
  \centerline{\epsfbox{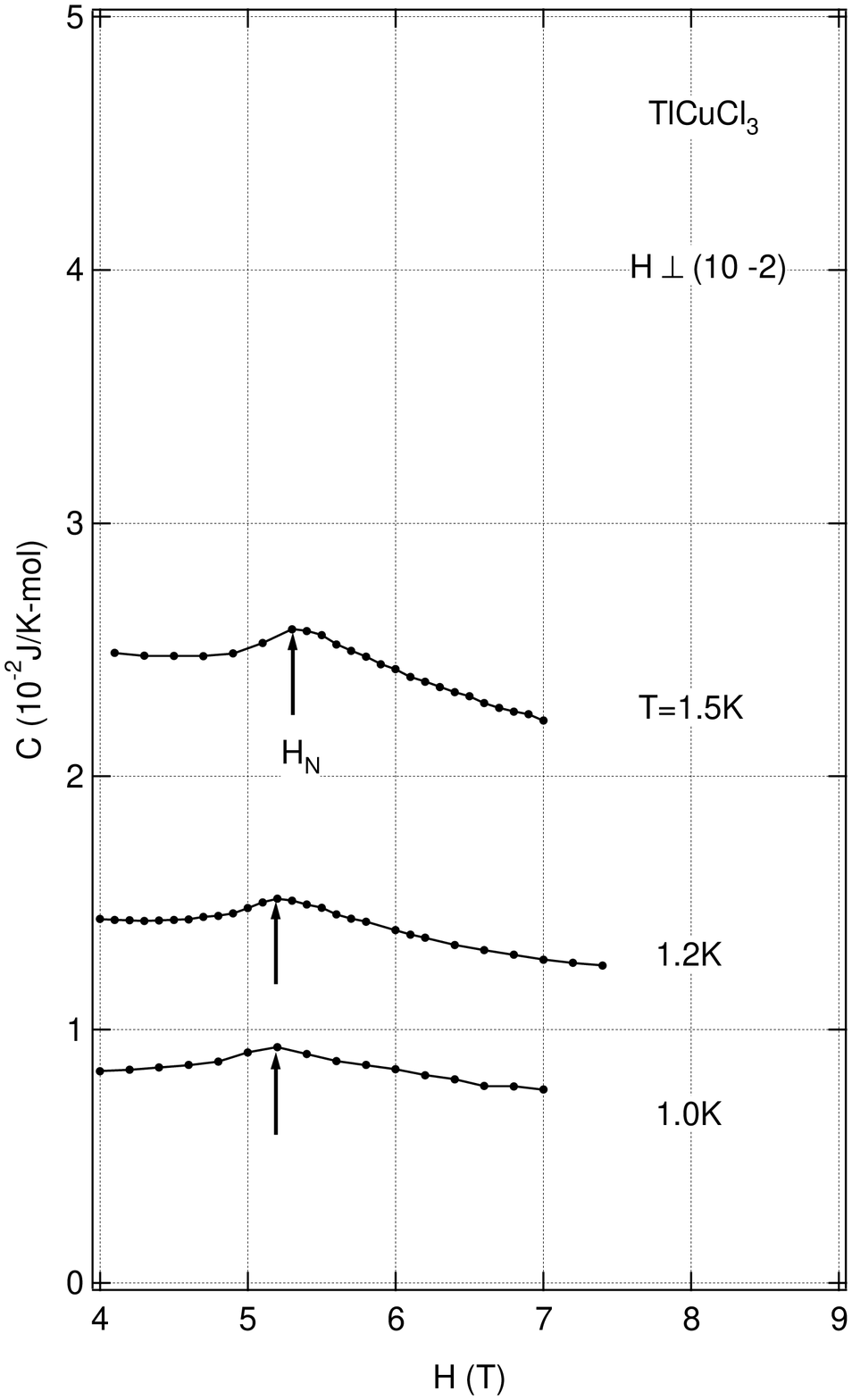}}
\begin{center}
(b)
\end{center}
\end{minipage}
\vspace*{1cm}
	\caption{The field dependence of the specific heat in TlCuCl$_3$ at various magnetic fields for (a) $H\parallel b$ and (b) $H\perp (1,0,{\bar 2})$.}
	\label{Hscan}
\end{figure}

In order to determine the transition point at a temperature lower than 3 K, we measured the field dependence of the specific heat at various temperatures. Examples of the measurements for $H \parallel b$ and $H \perp (1,0,{\bar 2})$ are shown in Fig. \ref{Hscan}. We assign the field at which the specific heat exhibits a cusplike anomaly to the transition field $H_{\rm N}(T)$. \par
\begin{figure}[ht]
\vspace*{5cm}
\begin{center}
 \epsfxsize=120mm
  \centerline{\epsfbox{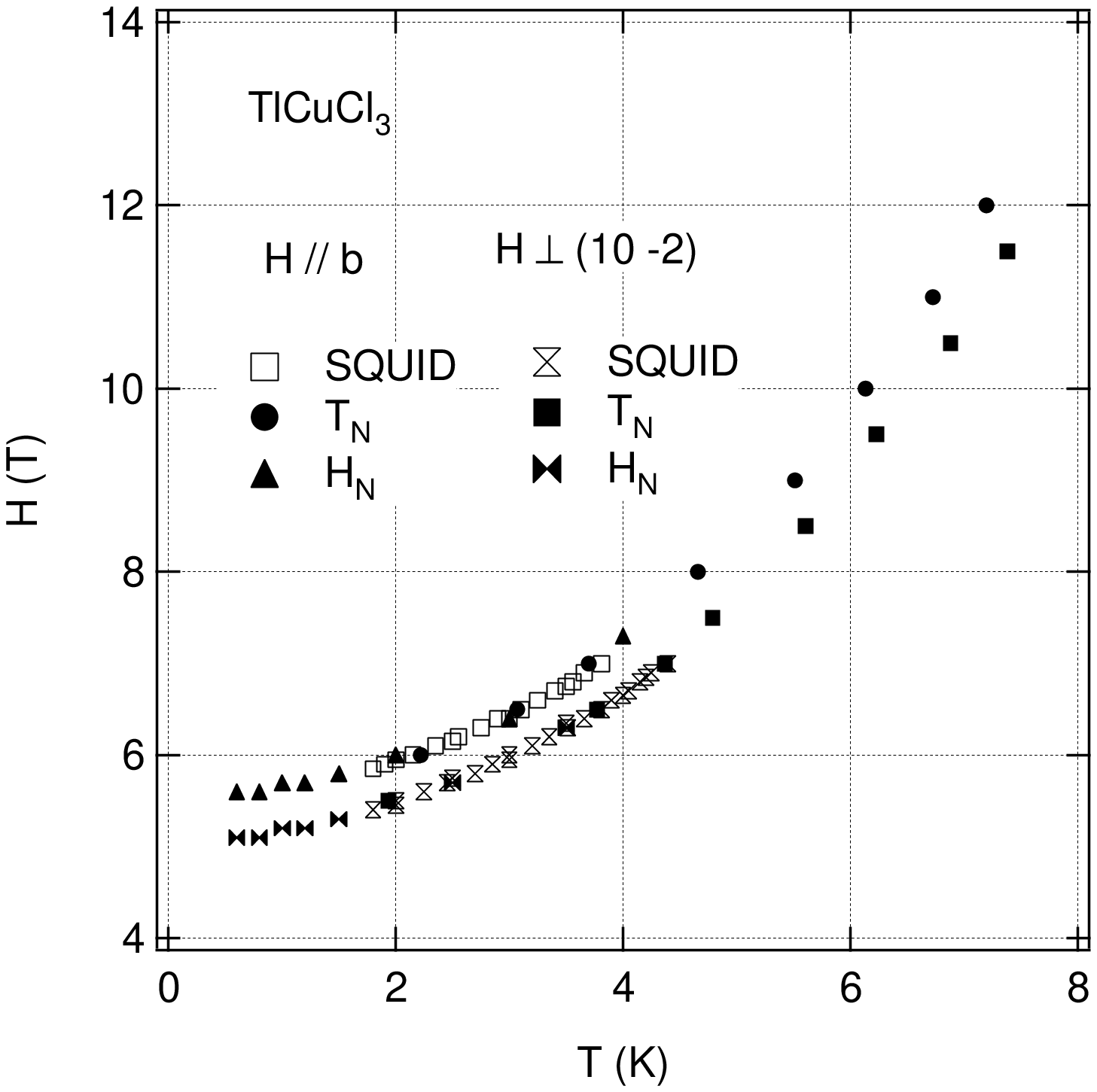}}
\end{center}
	\caption{The phase boundaries in TlCuCl$_3$ for $H\parallel b$ and $H\perp (1,0,{\bar 2})$. }
	\label{phase}
\end{figure}

The phase transition data obtained by temperature and field scans are summarized in Fig. \ref{phase}. In this figure, the transition points obtained from previous magnetization measurements are also plotted. The transition points determined from the present specific heat measurements and the magnetization measurements fall on the same line for $H \parallel b$ and $H \perp (1,0,{\bar 2})$. This confirms that the phase transition occurs at a temperature at which the magnetization has the minimum \cite{minimum}, as predicted based on the magnon BEC theory \cite{Nikuni}. The magnetization behavior is described by the theory as follows. Since the magnon corresponds to the excited triplet, the magnetization is proportional to the number of magnons. With decreasing temperature, the number of thermally excited magnons decreases for $T>T_{\rm N}$, while for $T<T_{\rm N}$ the BEC of magnons occurs, where the increase of the number of condensed magnons is larger than the decrease of the number of noncondensed magnons, so that the total number of magnons increases. \par
\begin{figure}[ht]
\vspace*{5cm}
\begin{center}
 \epsfxsize=120mm
  \centerline{\epsfbox{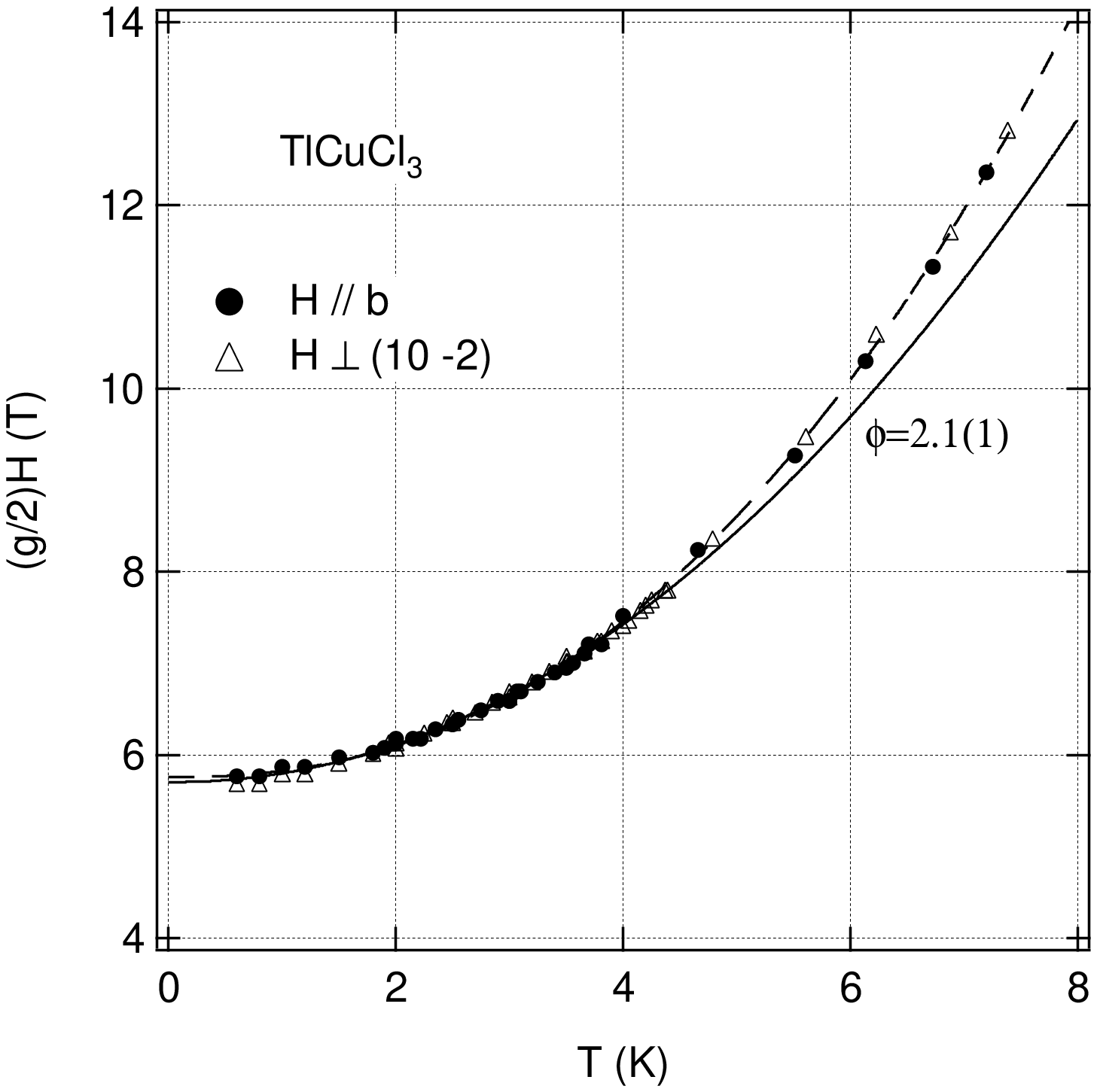}}
\end{center}
	\caption{The phase diagram in TlCuCl$_3$ normalized by the $g$-fator. The solid line denotes the fit with equation (1) with $\phi=2.1(1)$ and $gH_{\rm g}/2=5.7(1)$ T. The dashed line is a guide for the eyes.}
	\label{phasenor}
\end{figure}

Since the phase boundary depends on the $g$-factor, we normalize the phase diagram using the $g$-factor. The $g$-factors obtained by ESR measurements are $g=2.06$ for $H \parallel b$ and $g=2.23$ for $H \perp (1,0,{\bar 2})$. Figure \ref{phasenor} shows the phase diagram normalized by the $g$-factor. The phase boundaries for $H \parallel b$ and $H \perp (1,0,{\bar 2})$ coincide down to 0.6 K. This reconfirms that the phase boundary is independent of the external field direction when normalized by the $g$-factor, and that the anisotropy is negligible in TlCuCl$_3$. \par
The phase boundary can be described by the power law of eq. (1), as predicted based on the magnon BEC theory \cite{Nikuni}. We fit eq. (1) to the data for $T<4$ K, which is lower than half the gap temperature $\Delta/k_{\rm B}=7.5$ K. The solid line in Fig. \ref{phasenor} is the fit with $\phi=2.1(1)$ where $gH_{\rm g}/2=5.7(1)$ T.
The exponent $\phi=2.1$ obtained from the present measurements is somewhat larger than the value $\phi=3/2$ predicted by the theory \cite{Nikuni}. Since the theory is based on the Hartree-Fock approximation, the disagreement between the experimental and theoretical values of $\phi$ may be attributed to the fluctuation effect. \par


We have presented the results of specific heat measurements on the spin gap system TlCuCl$_3$. The field-induced 3D magnetic ordering was clearly observed. The phase boundary between the paramagnetic phase and the 3D ordered phase was determined, as shown in Fig. \ref{phasenor}. The present results confirm that the field-induced magnetic ordering takes place at the temperature where the magnetization has the cusplike minimum, and that the phase boundary can be described by the power law near the critical field $H_{\rm g}$, as predicted based on the magnon BEC theory \cite{Nikuni}. Therefore we conclude that the field-induced 3D magnetic ordering in TlCuCl$_3$ can be captured by the magnon BEC picture. 


The authors would like to thank K. Katsumata for access to the Mag Lab$^{\rm HC}$ calorimeter system. A. O. was supported by the Research Fellowships of the Japan Society for the Promotion of Science for Young Scientists.

\newpage

\newpage

\newpage

\newpage

\begin{references}
\bibitem{minimum} In reference 9, we assigned the temperature at which the magnetization has an inflection point as $T_{\rm N}$. Later, in reference 19, we assigned the temperature at which the magnetization has the cusplike minimum as $T_{\rm N}$, as concluded based on the magnon BEC theory.
\end{references}
\end{document}